\documentclass[a4paper]{jpconf}
\usepackage{graphicx}

\def\sqr#1#2{{\vcenter{\hrule height.#2pt\hbox{\vrule width.#2pt height#1pt
\kern#1pt \vrule width.#2pt}\hrule height.#2pt}}}
\def\square{\mathchoice\sqr64\sqr64\sqr{4.2}3\sqr{3.0}3}

\begin{document}

\title[Large-scale structure formation in the framework of scalar-tensor theories]{Power spectrum of 
large-scale structure cosmological models  in the framework of scalar-tensor theories}
\author{M A Rodr\'\i guez-Meza}
\address{(Instituto Avanzado de Cosmolog\'\i a, IAC)}
\address{Instituto Nacional de Investigaciones
Nucleares, Col. Escand\'on, Apdo. Postal 18-1027, 11801 M\'{e}xico D.F., M\'exico}
\ead{marioalberto.rodriguez@inin.gob.mx; http://astro.inin.mx/mar}

\begin{abstract}
We study the large-scale structure formation in the Universe in the
frame of scalar-tensor theories as an alternative to general relativity.
We review briefly
the Newtonian limit 
of non-minimally coupled scalar-tensor theories 
and the evolution equations of the $N$-body system
that is appropriate to
study large-scale structure formation in the Universe. 
We compute the power-spectrum of the universe at present epoch
and show how the large-scale
structure depends on the scalar field contribution.
\end{abstract}

\section{Introduction}
During the last two decades high precision observations have established the existence of two
component of the universe. These are known as dark energy and dark matter. What are their
fundamental nature? this one of the questions we need to answer\cite{Breton2004}. 
One of the most succesful models to explain observations is the $\Lambda$CDM but it still has
some problems.

In the past we have proposed a dark matter model in the framework of the Newtonian limit
of a scalar-tensor theory\cite{mar2004}.
We have done $N$-body simulations using a numerical code developed to take into account
the scalar field (SF) force contribution which is of Yukawa type in its Newtonian limit and we have use it
to study several situations, like collision of protogalaxies\cite{mar2001}, dynamics and 
collision of galaxies\cite{Gabbasov2006,mar2007a},
and to study cosmological simulations\cite{mar2007b,mar2008,mar2009b}.
Recently we have used it to address a possible solution to two of the main problems the 
$\Lambda$CDM has, i.e., the tendency to over populate the formed halos of galaxies with satellites 
and the flatness 
problem\cite{ceronu2007c,ceronu2009}. 

Several authors have also been using similar dark matter models to
study large-scale structure formation,
see for instance\cite{Hellwing, Brandao, Nusser}.

One of the tools to characterize the clustering properties of a universe model is the power spectrum.
Constraints on cosmological models and on
cosmological parameters have been derived from measuring power spectrum. 
Therefore,
in this work
we consider a dark matter model build from the Newtonian limit of a 
scalar-tensor theory\cite{mar2004} and
use it to study the large-scale structure formation of the universe by analyzing 
the power-spectrum versus the parameters of the model.

\section{Evolution equations for a $\Lambda$CMD universe within the framework of 
a scalar-tensor theory}


Giving that we are interested in the study of the large-scale structure formation of the universe
we will use the Newtonian limit of the scalar-tensor theory. This limit applies in a region small
compared to the Hubble length $c H^{-1}$, with $c$ the speed of light and $H$ the hubble factor,
and large compared to the Schwarzschild radii of any collapsed objects.

In this case the Vlasov-Poisson equation in an expanding universe describes the evolution of the
six-dimensional, one-particle distribution function, $f(\mathbf{x},\mathbf{p})$\cite{Bertschinger1998}.
The Vlasov equation is,
\begin{equation}\label{Vlasov_eq}
\frac{\partial f}{\partial t} + \frac{\mathbf{p}}{m a^2} \cdot \frac{\partial f}{\partial \mathbf{x}} 
-  m \nabla \Phi_N (\mathbf{x}) \cdot \frac{\partial f}{\partial \mathbf{p}} = 0 
\end{equation}
where $\mathbf{x}$ is the comoving coordinate, 
$\mathbf{p}=m a^2 \dot{\mathbf{x}}$, $m$ is the particle mass, and $\Phi_N$ is
the self-consistent gravitational potential given by the
Poisson equation,
\begin{equation}\label{Poisson_eq}
\nabla^2 \Phi_N(\mathbf{x}) = 4 \pi G_N \, a^2 
\left[
\rho(\mathbf{x}) - \rho_b(t)]
\right]
\end{equation}
where $\rho_b$ is the background mass density. 
Eqs. (\ref{Vlasov_eq}) and (\ref{Poisson_eq}) form the Vlasov-Poisson equation,
constitutes a collisionless, mean-field approximation to the evolution of the full
$N$-body distribution. An $N$-body code attempts to solve 
Eqs. (\ref{Vlasov_eq}) and (\ref{Poisson_eq}) by representing the one-particle
distribution function as
\begin{equation}\label{discrete_distribution_eq}
f(\mathbf{x},\mathbf{p}) = \sum_{i=1}^N \delta(\mathbf{x}-\mathbf{x}_i)\, 
\delta(\mathbf{p}-\mathbf{p}_i)
\end{equation}
Substitution of (\ref{discrete_distribution_eq}) in the Vlasov-Poisson system of 
equations yields the exact Newton's equations for a system of $N$ gravitating
particles.

In the Newtonian limit of STT of gravity, 
the Newtonian motion equation  for a particle $i$ is written as\cite{mar2008}
\begin{equation} \label{eq_motion}
\ddot{\mathbf{x}}_i + 2\, H \, \mathbf{x}_i = 
-\frac{1}{a^3} \frac{G_N}{1+\alpha} \sum_{j\ne i} \frac{m_j (\mathbf{x}_i-\mathbf{x}_j)}
{|\mathbf{x}_i-\mathbf{x}_j|^3} \; F_{SF}(|\mathbf{x}_i-\mathbf{x}_j|,\alpha,\lambda)
\end{equation}
where the sum includes all  
periodic images of particle $j$,  and $F_{SF}(r,\alpha,\lambda)$ is
\begin{equation} \label{eq_F}
F_{SF}(r,\alpha,\lambda) = 1+\alpha \, \left( 1+\frac{r}{\lambda} \right)\, e^{-r/\lambda}
\end{equation}
which,  for small distances compared to $\lambda$,  is 
$F_{SF}(r\ll \lambda,\alpha,\lambda) \approx 1+\alpha \, \left( 1+\frac{r}{\lambda} \right)$ and, 
for long 
distances, is  $F_{SF}(r\gg \lambda,\alpha,\lambda) \approx 1$, as in Newtonian physics. 

To simulate cosmological systems,  the expansion of the universe has to be
taken into account.
Also, to determine the nature of the cosmological model we need to determine
 the composition of the
universe, i. e., we need to give the values of $\Omega_i$ for each component $i$, 
taking into account
in this way all forms of energy densities that exist at present.

In this work we will consider a model with only two energy density contribution.
One which is a pressureless and
nonbaryonic dark matter  with $\Omega_{DM} \approx 0.3$ that does not couple with radiation.
Other, that will be a cosmological constant contribution $\Omega_\Lambda \approx 0.7$. 
Also, the model must be consistent with a static SF according to the Newtonian limit we are using
in this work.
Thus, the scale factor, $a(t)$,  is given by the following Friedman model\cite{mar2009b},
\begin{equation} \label{new_friedman}
a^3 H^2= H_{0}^{2} \left[\frac{\Omega_{DM0} +  \Omega_{\Lambda 0} \, a^3}{1+\alpha} 
+  \left(1-\frac{\Omega_{DM 0}+\Omega_{\Lambda 0}}{1+\alpha} \right) \, a  \right]
\end{equation}
where $H=\dot{a}/a$,  $\Omega_{DM0}$ and $\Omega_{\Lambda 0}$ 
are the matter and energy density evaluated at present, respectively.  We have also demanded
that the universe be a flat one.
 To be consistent 
with the CMB  spectrum and structure formation numerical 
experiments, cosmological constraints must be applied on $\alpha$ in order for it to 
be within the range $(-1,1)$ \cite{Nagata2002,Nagata2003,Shirata2005,Umezu2005}.  

\section{Results}
In this section, we present results of cosmological simulations of a $\Lambda$CDM universe
with and without SF contribution. 
We start our simulations with an initial distribution of $N=2\times 32^3$ particles 
in a box with sides of $L=50 h^{-1}$ Mpc at $z=10$. 
This case is similar to the one that comes with Gadget 1 \cite{Springel2001}.
At present epoch, $\Omega_{m0} = 0.3$, $\Omega_{\Lambda 0}=0.7$,
$H_0= 100 h$ km/s/Mpc, $h=0.7$. We restrict the values of $\alpha$ to the interval $(-1,1)$ 
  \cite{Nagata2002,Nagata2003,Shirata2005,Umezu2005}  and  use $\lambda=5$ Mpc, since 
this scale turns out to be an intermediate scale between the size of the clump groups and the separation of 
the formed groups. However, we have analyzed elsewhere several values of $\lambda$\cite{mar2009b}.
Other universe models have been analyzed in \cite{mar2007b,mar2008,mar2009a}. 

Because the visible component is the smaller one and given our interest to
test the consequences of including a SF contribution to the evolution equations,
our model excludes gas particles, but all its mass has been added to the dark matter. 
\begin{figure}
\begin{minipage}{6.5in}
\begin{center}
\includegraphics[width=4.in]{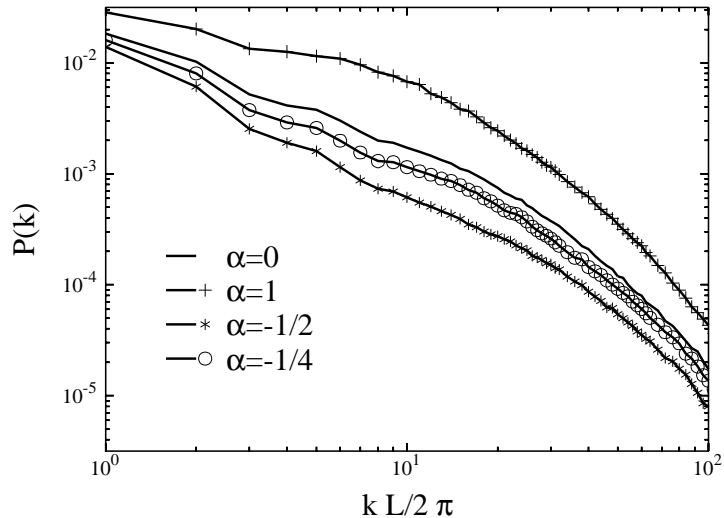}
\end{center}
\end{minipage}
\caption{Power spectrum of our scalar field universe model without and with SF.
See text for details. }
\end{figure}
In Fig. 1 we show the power spectrum at redshift $z=0$ of our  $\Lambda$CDM model for several
values of $\alpha$. The line without symbols shows the case of a  $\Lambda$CDM model 
without the contribution of the scalar field. The line with ($+$) is the case of a  $\Lambda$CDM model 
with scalar field with $\alpha=1$, the line with ($*$) uses $\alpha=-1/2$, and the line with (o) uses
$\alpha=-1/4$.
One notes clearly how the SF modifies the clustering of matter of the system. 
The most
dramatic cases are the models with $\alpha=-1/2$ and $\alpha=1$.

We now analyze the general effect the parameter $\alpha$ has on the clustering.  
For a given $\lambda$ the role of $\alpha$ in our approach can be seen in equations
(\ref{eq_motion})-(\ref{eq_F}) and is contained in the factor $F_{SF}(r,\alpha,\lambda)$.
The factor $F_{SF}$ augments (diminishes) for positive (negative)  
values of $\alpha$ for small distances compared to  $\lambda$, resulting in more (less) structure formation for positive (negative) values of $\alpha$ compared to the $\Lambda$CDM model.  
In the case of the line with ($+$), for $r \ll \lambda$,
the effective gravitational pull has been  augmented by a factor of $2$, 
in contrast to case shown with line and ($*$) where it has diminished  by a factor of 1/2; in the
model with the line and (o) the pull 
diminishes only by a factor of 3/4. That is why one observes for $r < \lambda$ more structure 
formation in case with line and ($+$), less in case with line and ($*$), and lesser in  model with line and (o).  
The effect is  then, for a growing positive $\alpha$, 
to speed up  the growth of perturbations, then of halos and then of clusters, whereas negative 
$\alpha$ values ($\alpha \rightarrow -1$) tend to slow down the growth.

\section{Conclusions}
The general gravitational effect of the scalar-tensor dark matter model is that the interaction changes
by a factor $F_{SF}(r,\alpha,\lambda)\approx 1 + \alpha(1+r/\lambda)$ for $r < \lambda$ in
comparison with the Newtonian case. For negative values of $\alpha$ the effect is to diminish
the formation of structures and the opposite ocurrs for positive values of $\alpha$. 
For $r>\lambda$ the dynamics is essentially Newtonian. By computing
the power spectrum we were able to determine the effects of the scalar field contribution to
the galaxy clustering. Of course we will need to do a systematic and with greater resolution
study and make detailed comparisons with observations to support or to rule out a dark
model based on scalar-tensor theories.


\ack{This work was supported by CONACYT, grant numbers CB-2007-84133-F, I0101/131/07 C-234/07, IAC collaboration. 
The simulations were performed in the UNAM HP cluster {\it Kan-Balam}.}

\end{document}